# Brains and language models converge on a shared conceptual space across different languages

Zaid Zada[1*], Samuel A. Nastase[1], Jixing Li[2], Uri Hasson[1]

[1] Department of Psychology and Neuroscience Institute, Princeton University, New Jersey, 08544, US.
[2] Department of Linguistics and Translation, City University of Hong Kong
* Corresponding author. Email: zzada@princeton.edu

## Abstract

Human languages differ widely in their forms, each having distinct sounds, scripts, and syntax. Yet, they can all convey similar meaning. Do different languages converge on a shared neural substrate for conceptual meaning? We used language models (LMs) and naturalistic fMRI to identify neural representations of the shared conceptual meaning of the same story as heard by native speakers of three languages: English, Chinese, and French. We found that LMs trained on entirely different languages converge onto a similar embedding space, especially in the middle layers. We then aimed to find if a similar shared space exists in the brains of different native speakers of the three languages. We trained voxelwise encoding models that align the LM embeddings with neural responses from one group of subjects speaking a single language. We then used the encoding models trained on one language to predict the neural activity in listeners of other languages. We found that models trained to predict neural activity for one language generalize to different subjects listening to the same content in a different language, across high-level language and default-mode regions. Our results suggest that the neural representations of meaning underlying different languages are shared across speakers of various languages, and that LMs trained on different languages converge on this shared meaning. These findings suggest that, despite the diversity of languages, shared meaning emerges from our interactions with one another and our shared world.

## Introduction

With over 7,000 languages worldwide, human languages are vastly diverse, encompassing a wide range of sounds, alphabets, syntax, and more. For almost any seemingly "universal" rule of language, there exist countervailing examples among the world's languages (Evans & Levinson, 2001). That said, while the forms of language differ, their content and function must be shared, to some extent, across different languages. For example, regardless of how any one language refers to a mother or father, these concepts are shared across many communities, and can be translated between languages. All languages enable us to convey similar concepts through our shared experiences and interactions with the world. Can we access the shared conceptual meaning space behind the incredible diversity of linguistic forms? How is shared meaning across languages encoded in the brain?

Some clues come from studies of language processing across different modalities. Neuroscientists have identified similar brain regions associated with over 40 spoken languages (Malik-Moraleda et al., 2024) and sign languages (MacSweeney et al., 2008; Neville et al., 1998). Brain regions supporting language comprehension are engaged by both listening and reading (Regev et al., 2013; Deniz et al., 2019), as well as speaking and listening (Stephens et al., 2014; Zada et al., 2024, 2025). Studies of bilingual speakers have found similar conceptual representations for different languages using isolated words or sentences (Abutalebi & Green, 2007; Buchweitz et al., 2011; Correia et al., 2014; Zinszer et al., 2016), as well as spoken narratives (Chen et al., 2024b). However, localizing the processing of different languages to overlapping brain regions alone cannot speak to the shared content encoded in brain activity across languages. Bilingual speakers comprehend both languages and can generally translate between them, suggesting that their conceptual understanding is interconnected rather than isolated.

In a critical step forward, Honey and colleagues (2012) directly tested for cross-language alignment in English and Russian speakers. Both sets of participants listened to the same story in English and in Russian during fMRI scanning. First, Honey and colleagues localized neural activity related to the processing of speech sounds, independent of comprehension, by correlating the neural responses of English speakers (who did not understand Russian) with those of Russian speakers, while both groups listened to the story narrated in Russian. They found neural alignment driven by speech sounds, regardless of comprehension, only in the auditory cortex. Next, the researchers focused on the neural activity associated with the processing of shared meaning across languages, regardless of the speech sounds used. They did this by comparing the brain activity of English speakers listening to the story narrated in English with that of Russian speakers listening to the same story narrated in Russian. In this case, both the speech sounds and certain linguistic properties (e.g., morphological and syntactic structures) differ between the two languages, but the overall conceptual meaning of the narrative is shared. The results revealed neural alignment linked to shared meaning in the temporal, frontal, and medial parietal regions of the brain, demonstrating that this alignment occurs independently of the language being spoken. Building on these ideas, we aim to pursue a theory for how different languages may converge in a neural population code. Specifically, we contend that different languages are encoded in a multidimensional neural embedding space, that the meaning of a given language is encoded in the geometric structure of its respective embedding space, and that these geometries are largely preserved across different languages.If this is the case, we should be able to explicitly model these shared features in neural activity across speakers of different languages.

Modern neural networks for language processing—language models (LMs)—can serve as explicit computational models of natural language processing in the human brain. They learn distributed representations of words that reside in a structured, high-dimensional embedding space, where different structures of language are encoded in the geometric structure of the embedding space (Manning et al., 2020). Prior work has shown that the embedding space of a unilingual LM (i.e., trained on only one language) can be rotated to align with another embedding space of an LM trained on a different language, allowing it to translate novel words (Mikolov et al., 2013; Conneau et al., 2018; Fig. 1A). Newer LMs are becoming increasingly multilingual. Multilingual BERT, for example, was trained on over 100 languages (Devlin et al., 2018), forcing the model to embed shared multilingual features in a single, unified high-dimensional embedding space (Chi et al., 2020; Chang et al., 2022; Stańczak et al., 2022; de Varda & Marelli, 2023). Multimodal language models, such as Whisper (Radford et al., 2022) and SeamlessM4T (SEAMLESS Communication Team et al., 2025), can transcribe, translate, and speak over 100 languages. Recent work on interpreting multilingual language models has found that multilingual conceptual features are encoded at intermediate layers of the neural network (more so than, e.g., shared syntactic structure), and that language-specific features are relegated to early and late layers (Lindsey et al., 2025). Recent studies have shown that LM embeddings better align with human brain activity than other kinds of language representations (Schrimpf et al., 2021; Caucheteux & King, 2022; Goldstein et al., 2022). However, most studies to date have only analyzed language processing in a single language (typically English) and with unilingual models (cf. Nakagi & Matsuyama, 2025; de Varda et al., 2025).

Here, we used a single, 1.5-hour-long spoken story narrated in three different languages (English, French, and Chinese) to identify the extent of shared cross-language (supralingual) features in language models and human brain activity. Three groups of participants listened to the story in their native language during functional magnetic resonance imaging (fMRI). We hypothesized that the brain activity for listeners should be aligned by the high-level conceptual meaning of the story across different languages (like Honey and colleagues, 2012). We next aimed to explicitly model the content driving this alignment across languages: we hypothesized that supralingual alignment can be modeled using the embedding spaces derived from three different LMs, each trained on a single language (Fig. 1B). We first confirmed that the word embeddings for the three different language transcripts converge on a shared geometry, especially in the middle layers (in both unilingual and multilingual language models). We next used voxelwise encoding models to map LM word embeddings onto each subject's brain activity and tested how well the models generalize across subjects and to different languages. We found that both unilingual and multilingual LM embeddings mapped onto brain activity across subjects within their respective languages. Crucially, we also found that encoding models mapping from LM embeddings onto brain activity in one language (e.g., English embeddings → English listener's brain) also generalized to Chinese and French speakers listening to the same story in their respective languages. Finally, using a multimodal speech-to-text model, we discovered shared multilingual features of neural activity captured by both speech (sound) and language (word) embeddings.

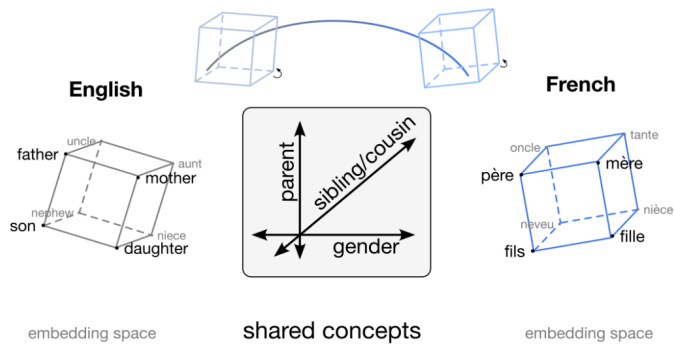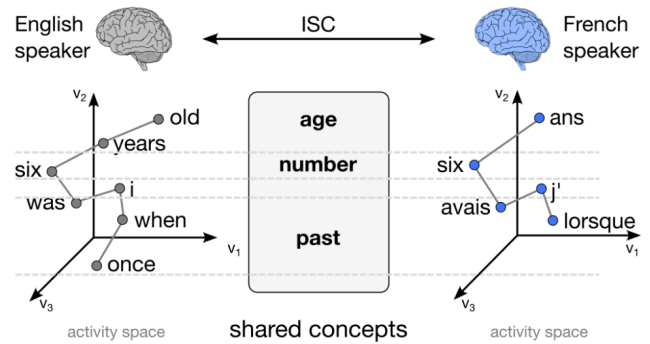

**Fig. 1. Schematic hypothesis of convergence on concepts across languages.** (**A**) Neural network models for language learn to embed words in a high-dimensional space with meaningful geometric structure, as illustrated by the kinship example from Rumelhart and Abrahamson (1973). The structure is preserved across languages and can be aligned by a simple rotation. (**B**) The neural activity of English (gray) and French (blue) speakers is correlated when they listen to the same content in their respective languages. The sentence forms a trajectory in activity space, where concepts are shared but forms are different.

## Results

How do different languages converge on a shared neural substrate for conceptual meaning? We utilized an open fMRI dataset in which three groups of participants listened to an audiobook of The Little Prince in their respective native languages: English (n = 49), Mandarin Chinese (n = 35), and French (n = 29; Li et al., 2022). The audiobook was ~100 minutes long and was divided into 9 scanning sessions. We aligned the audio and text across the three languages for each sentence, resulting in a total of ~1,650 sentences (Fig. 2A). Using the transcripts, we extracted contextual word embeddings from three types of language models: (1) three *unilingual* BERT models: one trained solely on English text, one trained solely on Chinese text, and one trained solely on French text (Devlin et al., 2018); (2) a *multilingual* BERT model trained on Wikipedia corpora of 100 languages; and (3) a multilingual speech transcription and translation language model, Whisper (Radford et al., 2022; Fig. 2B).

We trained voxelwise encoding models to predict each subject's BOLD responses from the word embeddings for their native language to quantify the linguistic and conceptual meaning of the story (Fig. 2C). On a held-out run, we evaluated encoding performance by correlating the model-predicted BOLD response for each subject with the actual BOLD time series averaged across other subjects. Critically, we evaluated how well the encoding model predictions generalize across languages. For example, we evaluated predictions from the English model trained on the brain activity of an English subject against the brain activity of French and Chinese subjects (Fig. 2D). The timing of words and sentences differed across the three language audiobooks. For simplicity, when evaluating across languages, we first downsampled the TR-level time series into the sentence level by averaging the activity of all TRs within each sentence. We fit encoding models in a cross-validation paradigm: training on eight runs, and evaluating accuracy on the held-out ninth run across the three language groups; encoding results were then averaged across the left-out test runs.

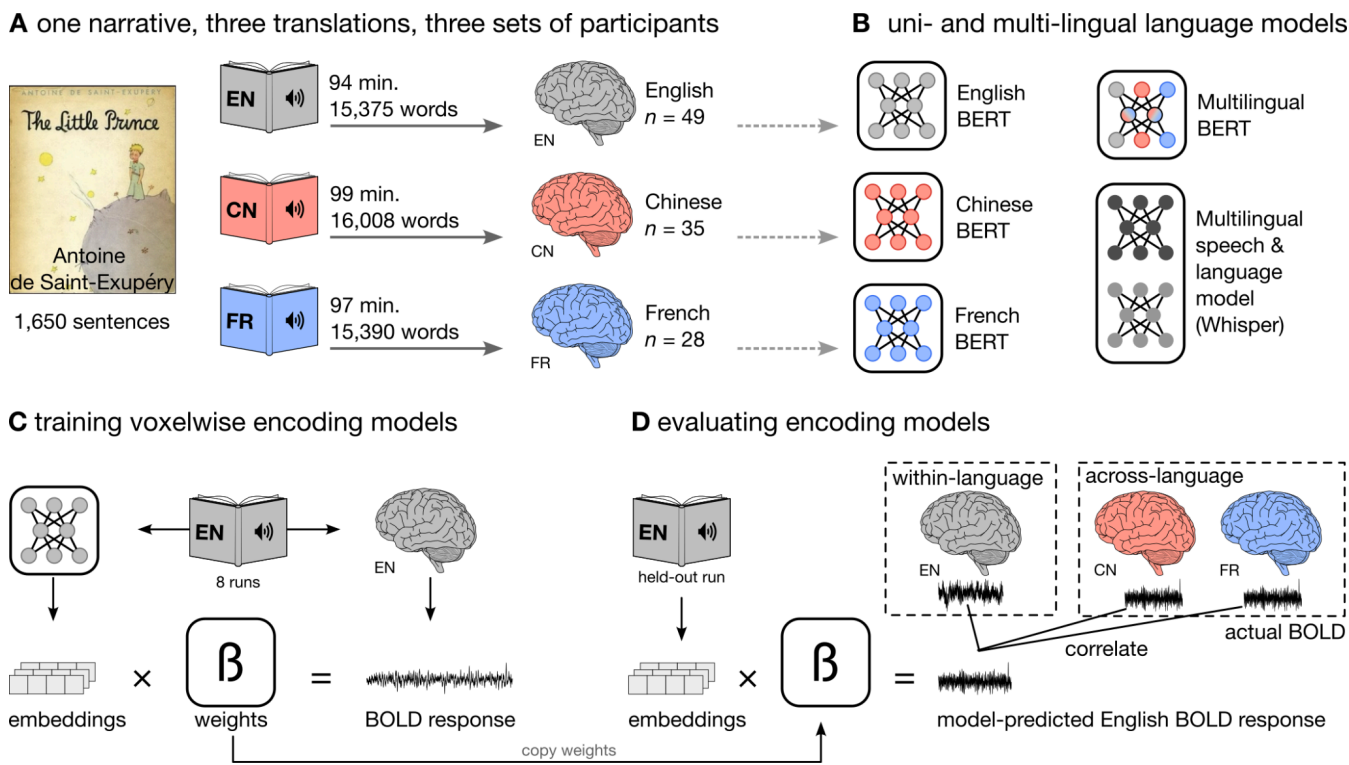

**Fig. 2. Dataset description and modeling framework.** (**A**) We used an open fMRI dataset where three groups of participants listened to an audiobook of the same story in their native language: English (gray), Chinese (red), or French (blue). (**B**) We provided the story transcripts as input to language models that are either unilingual (trained on a single language), multilingual (trained on multiple languages), or multimodal (combining speech and language across multiple languages). (**C**) We extracted contextual word embeddings from each unilingual language model using its "native" language. Then, we trained voxel-wise encoding models to predict a subject's BOLD responses from their corresponding language model. (**D**) On a held-out test set, we generated model-predicted BOLD responses and evaluated them against the average time series for each language group.

*Unilingual models and brains converge on shared linguistic representations*

We hypothesized that language models trained on different language corpora would nonetheless converge on similar conceptual representations. To test this, we computed the similarity of unilingual BERT (uBERT) sentence embeddings across three pairs of languages. We derived sentence embeddings by averaging the token-level embeddings within each sentence. The BERT models shared the same underlying architecture and a similar number of overall parameters. The main difference is that they were each trained on a separate, and only one, language corpus in English, French, or Chinese. Because their learned embedding spaces are oriented in arbitrary directions, we aligned them using a rotation-only linear transformation. Specifically, we learned a rotation between a pair of language embeddings on half of all sentences ($n = 825$). Then we evaluated the rotation on the remaining sentences by computing the correlation between the reference language embeddings and the rotated language embeddings. We found notable similarities (average $r$ across all test words and layers = 0.115) between each pair of language embeddings (Fig. 3A). The language pair similarities increased along the model depth from the first layer ($r = 0.054$) to the late-intermediate layers ($r = 0.154$), with the last layer exhibiting a drop in similarity ($r = 0.116$). We also observed that English and

French were generally more similar than French–Chinese and English–Chinese. These correlations are relative to a control baseline using embeddings from *untrained* BERT models (see Fig. S1 for raw correlations). These results suggest that language models trained on only one language nonetheless learn an embedding space that shares an internal structure with other languages. These unilingual models converge on this shared geometry only in service of better predicting words in their respective languages.

We next used the uBERT word embeddings to identify neural representations of the shared conceptual meaning of the same story as heard by the speakers of three different languages. We trained encoding models to predict each subject's BOLD activity using their native language's story and the BERT model trained on that language (Fig. 2C). First, we evaluated encoding performance within each language group by correlating a subject's model-predicted time series with the average BOLD time series for their own group (for a left-out segment of the stimulus). We found that word embeddings corresponding to a subject's native language significantly predicted voxels across a broad network of cortical areas associated with language and narrative comprehension: superior temporal cortex, frontal cortex, parietal regions, and medial default-mode regions (Fig. 3B).

Crucially, we next evaluated a subject's model-predicted time series against the other language groups (Fig. 2D). We found that unilingual word embeddings trained to predict native speakers of one language generalized to BOLD responses of subjects listening to the story in their own respective native language (Fig. 3C). For example, we found that encoding models trained to map English uBERT embeddings onto an English-speaking subject's brain activity while they listened to the English version of the story generalize well to the brain activity in both French and Chinese subjects listening to the story in French and Chinese. Most regions that had significant encoding performance within a speaker's language group were also well predicted in other language groups (correlation between within- and across-language unthresholded brain maps, $r = 0.974$). However, the overall encoding performance across language pairs was slightly weaker than within languages, with significant differences in early auditory cortex and superior temporal gyrus ($p_{FDR} < .05$; Fig. S2), which likely host language-specific speech processing. These results suggest that word embeddings can capture conceptual representations that are at least partially shared across brains when speakers of different languages listen to the same story, but in different languages. We found qualitatively similar results when training and testing encoding models at the level of individual words instead of sentences (Fig. S3). The magnitude and cortical extent of supralingual encoding suggest that much of what drives the alignment between LMs and brain activity is not specific to a particular language.

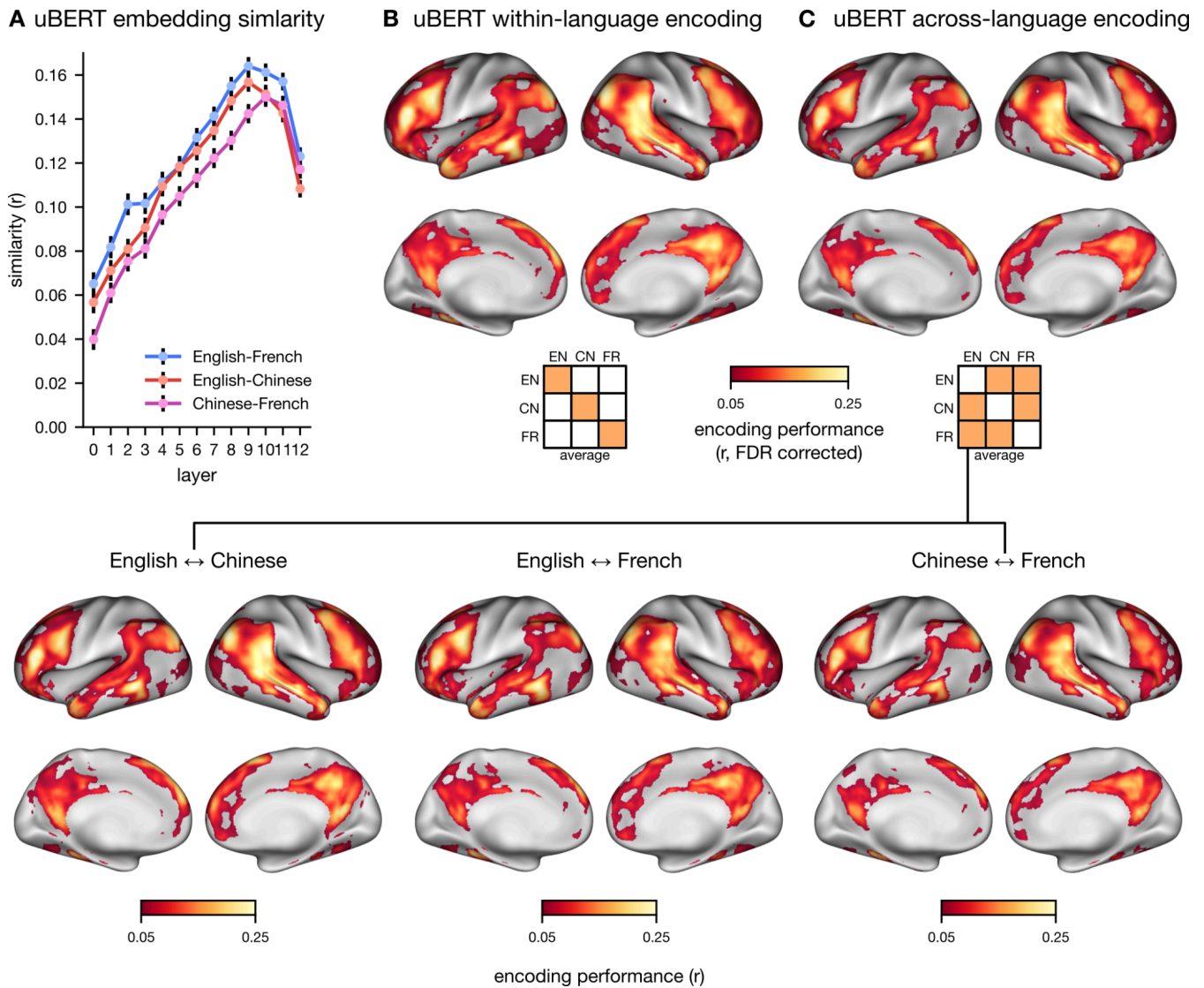

**Fig. 3. Shared content across different languages drives similarity between models and brains.**
(**A**) We compared sentence embeddings from three unilingual BERT (uBERT) models: English, Chinese, and French. (**B**) We fit encoding models using uBERT embeddings onto the brain activity of their language-congruent participants. We evaluated encoding model accuracy across subjects on a held-out segment of the stimulus within each language, then averaged across three languages. (**C**) We evaluated how well encoding models trained on one language generalize to another language. All maps are thresholded for statistically significant encoding performance (one-sided t-test, $p_{FDR} > .05$).

*Multilingual embeddings reflect the familial structure of language*
Language models are increasingly becoming multilingual, forcing words from different languages to coexist in one shared high-dimensional embedding space. Given the pressure to encode multiple languages in the same embedding space, we expected that word embeddings from multilingual language models would be more similar across our three languages than separately trained unilingual models. We extracted contextual word embeddings from the same model, multilingual BERT (mBERT), separately for the transcripts in each language. We computed the correlation between sentence embeddings (after averaging individual token embeddings per sentence) across the three language pairs. Unlike with the unilingual model comparison, we did not need to learn or apply a rotation to the embeddings because all embeddings are derived from a single model. We found that sentence similarity increases with model depth up to layer eight, then gradually decreases. This suggests that early and late layers of mBERT are more language-specific than middle layers, supporting recent results in LM interpretability (Lindsey et al., 2025). As before, we observed that English and French are most similar to one another, followed by English–Chinese and Chinese–French. These results suggest that language models trained on multiple languages leverage conceptual similarities across languages and learn features that are shared across multiple languages.

Next, we surveyed the landscape of shared features in mBERT's multilingual embedding space across 58 languages. We translated the English transcript into 55 languages that were both supported by mBERT and our translation tool (Table S1). All 58 languages (55 plus English, Chinese, and French) constituted 14 distinct language families, with Indo-European being the largest, comprising 36 languages. We extracted contextual word embeddings from mBERT's middle layer for each language transcript and then averaged the token embeddings into sentence embeddings. Next, we computed a 58 × 58 language-by-language representational dissimilarity matrix using correlation distance and projected it onto two dimensions using multidimensional scaling (MDS). We found clusters of languages grouped by their familial hierarchical structure. For example, Indo-European families were clustered into Germanic, Romance, Balto-Slavic, and Indo-Iranian groups (Fig. 4B). Dravidian languages were also more similar to each other than to other languages. Not all Indo-European languages represented a sparse set of families, with some having only one language per family (e.g., Japanese, Korean, Arabic). Interestingly, Mandarin Chinese was closest to Japanese and Vietnamese. These findings suggest that mBERT learns to embed languages that are more closely related nearer to each other in its high-dimensional embedding space.

Given that the story's translations appear to exhibit a coherent embedding structure, we hypothesized that languages more similar to our subjects' native languages would be better able to predict their neural activity. For example, an English participant would hear the sentence "Once, when I was six years old…", which has the same meaning in German, "Einmal, als ich sechs Jahre alt war…", and Arabic, "ذات مرة، عندما كنت في السادسة من عمري...". We anticipated that an encoding model using German embeddings from mBERT would outperform one using Arabic embeddings in predicting an English subject's BOLD response, as German is more closely related to English. For each subject, we trained an encoding model using their native language (for example, we used the English embeddings from mBERT for English speakers) across a subset of voxels with high sentence-level intersubject correlations (Fig. S4). However, during evaluation on the test set, we replaced the English embeddings with those from one of the 57 other languages. This approach allowed us to assess the degree of zero-shot generalization from the embedding space of one language to another in predicting brain activity. We found that languages with embeddings more similar to a subject's native language better predicted their brain activity (Fig. 4C). For example, embeddings for many Romance languages better predicted French listener's brain activity than non-Romance languages ($r = 0.786$, $p < 1e-10$). Similarly, Germanic language embeddings better predicted brain activity in native English speakers ($r = 0.869$, $p < 1e-10$). Chinese embeddings were most similar to those of Vietnamese and Japanese, but performed similarly to those of other languages. We suspect that our subset of languages, and perhaps the models themselves, are biased towards Indo-European languages, which may limit the accuracy of our interpretation of the results when applied to Chinese. Moreover, we translated the story using the English transcript as a source, which may have further biased the translation toward languages with similar structures. In sum, we found that languages more similar to the listener's native language, both in terms of language family and embedding similarity, better predicted their brain activity than language embeddings that were more dissimilar. This result reinforces the claim that linguistic features from different languages can be learned by a single language model and aligned into a unified, high-dimensional embedding space.

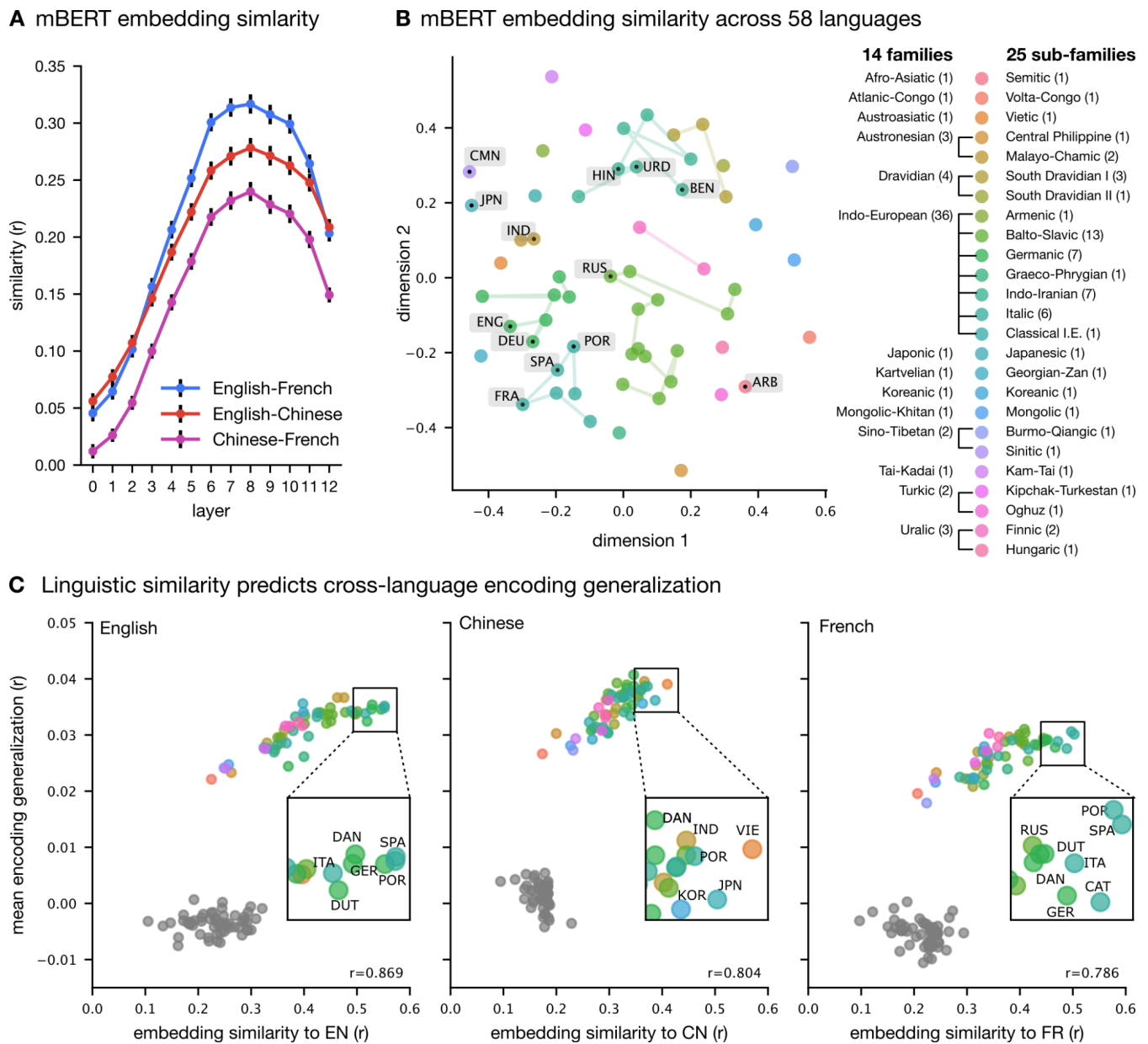

**Fig. 4. Language family structure reflected in multilingual language model embeddings. (A)** We compared sentence embeddings from multilingual BERT (mBERT) models (English, Chinese, and French). **(B)** We computed a pairwise dissimilarity matrix between mBERT embeddings across all 58 languages and projected it onto two dimensions using multidimensional scaling. Each circle marker is one language, colored according to its subfamily categorization. To orient the reader, we annotated

several languages with the largest number of speakers. Legend on the right denotes a simplified family hierarchy, with the number of languages within each (sub)family in parentheses. (**C**) We compared how similar each of the 58 languages' embeddings were to our three main languages (x-axis), and evaluated the encoding performance of the already trained model using each language's embeddings (y-axis). As a control, we repeated the same analysis for using untrained mBERT embeddings (gray dots; not included in the reported correlation).

***Shared speech features across languages***

So far, the language models we used have only had access to text transcripts, whereas the participants listened to the story (e.g., instead of reading it). Lastly, we tested whether any speech-specific features may be shared across languages and brains. To that end, we extracted speech and text embeddings from the multilingual speech-to-text language model, Whisper (Radford et al., 2023). Whisper employs an encoder-decoder transformer architecture, where raw audio spectrograms serve as input to the encoder, and the corresponding text tokens are used as input to the decoder. This allows the autoregressive decoder to attend to speech features from the last layer of the encoder as it transcribes. Using both the stimulus audio and transcript for each language (English, Chinese, and French), we extracted Whisper activations for each layer and aggregated them into sentence embeddings to compare across each pair of languages. We found that acoustic features in the encoder progressively become more similar and peak in the last layer (Fig. 5A). Conversely, in the decoder, cross-language embedding similarity peaks along the middle layers and decreases in the early and later layers, mirroring the trend in mBERT (Figs. 3A, 4A). We again noticed that English and French were more similar to each other than to Chinese. These findings suggest that a multilingual and multimodal language model can represent shared multilingual speech features, despite obvious differences in the acoustic patterns of different spoken languages.

How is it that abstracted speech representations can share features across widely different languages? Shared features may encode overlapping sets of phonemes or similar patterns of prosody or stress. To test this hypothesis, we identified the phonemes of every word in all three languages using a multilingual international phonetic alphabet (IPA) dictionary. We found 63 phonemes in total, 19 of which were shared across English, French, and Chinese. We then built linear probes to classify the presence of each of the 19 phonemes using embeddings from the encoder's first and last layers. We trained each probe on one language (e.g., English) using half the story to fit the model, and tested it on the other languages (e.g., French and Chinese) in the other half of the story. We evaluated the classifier using adjusted balanced accuracy, which sets the classifier's chance level at zero. We found that the last layer was more accurate in predicting phonemes in the other languages than the first layer (Fig. 5B). The English classifier was better able to predict French phonemes than Chinese, and vice versa. Overall, this suggests that Whisper's encoder leverages shared speech features across languages, which emerge particularly in the later layers of the encoder.

How do these speech and text embeddings map onto the brain? We fit voxelwise encoding models using both speech embeddings from Whisper's last encoder layer and word embeddings from Whisper's middle decoder layer. We trained the encoding models using banded ridge regression to ensure both types of embeddings can fairly compete for variance in neural activity. We evaluated the models across language groups, as we did in our previous analyses. We found that speech features predicted early auditory cortex (EAC), as well as some prefrontal and medial areas, across languages. In contrast, the higher-level word features broadly predicted typical language network regions as well as posterior medial cortex across languages (Fig. 5C). Many language regions exhibited mixed selectivity for both speech and word embedding features. We contrasted the two feature spaces by subtracting speech feature performance from word embedding performance. We found that speech embeddings more accurately predicted EAC, the anterior insula, and certain dorso-medial regions. The word embeddings, instead, better predicted inferior frontal and temporo-parietal cortex, as well as precuneus. These findings suggest that shared meaning across languages is not limited to abstract conceptual representations but that lower-level speech features are also shared.

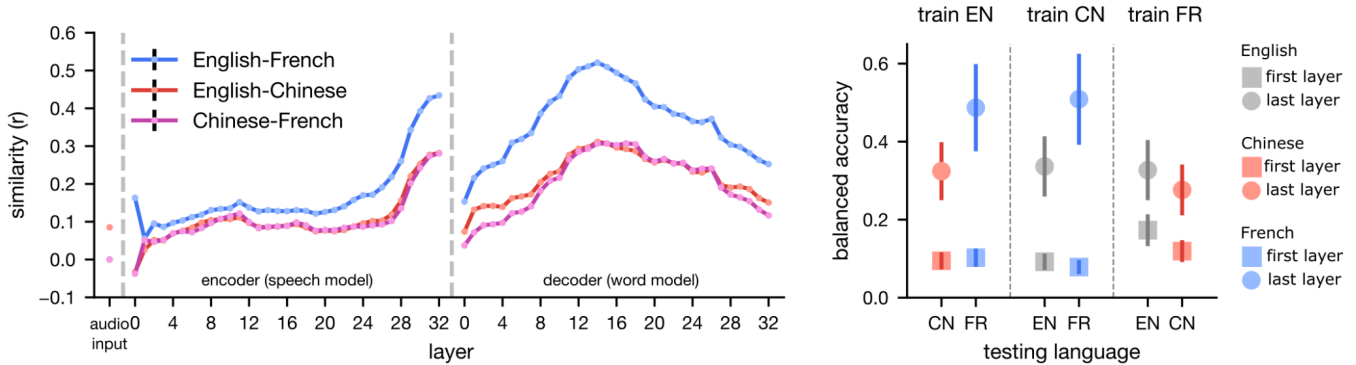

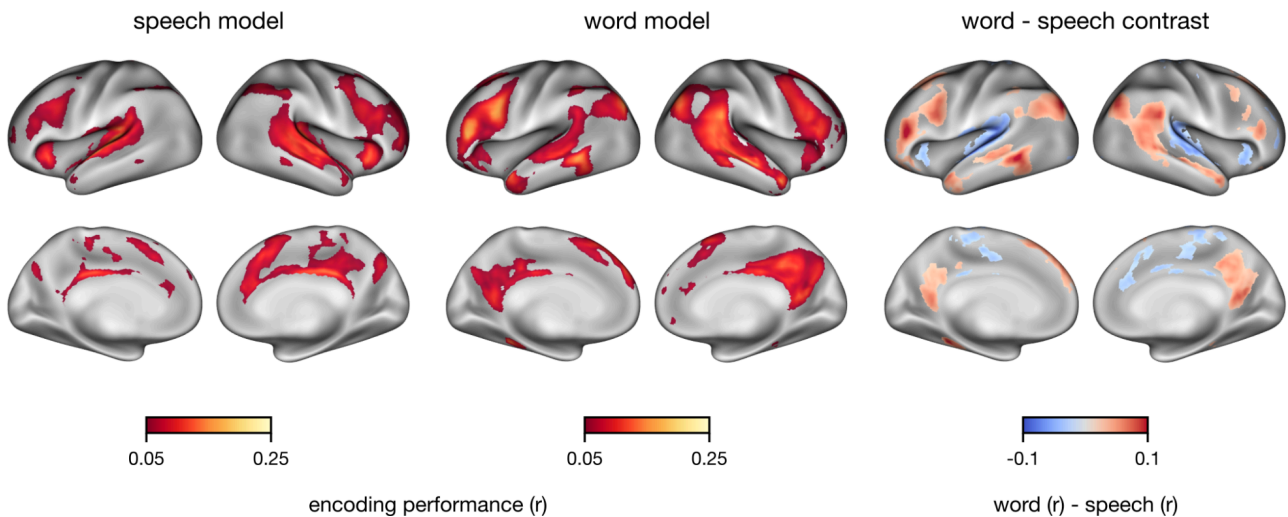

**Fig. 5. Shared multilingual speech features in Whisper and across brains.** (**A**) We compared sentence embeddings from the multimodal language model Whisper across languages. The encoder processes acoustic speech inputs; the decoder processes text inputs, but can "attend" to the last layer of the encoder. (**B**) We trained classifier probes to decode each phoneme from the embeddings of the first and last encoder layers. Each classifier was trained on one language's word embeddings and tested on the two other languages. (**C**) We fit voxelwise encoding models jointly with speech embeddings (the last encoder layer) and word embeddings (the middle layer of the decoder). We evaluated encoding models across languages and averaged performance across language pairs. We present the relative correlation of each feature space and the difference between them.

## Discussion

Our findings revealed that brain activity in native speakers of three different languages is aligned along conceptual features that are shared across languages. We first demonstrated that sentence embeddings converge in silico, in both unilingual (Fig. 3A) and multilingual (Fig. 4A) language models, as well as in speech-based multimodal language models (Fig. 5A). We then trained voxelwise encoding models to map LM embeddings to brain activity. Encoding performance was strongest in areas associated with language (superior temporal, inferior frontal, and parietal regions) and narrative comprehension (posterior medial cortex; Fig. 3B). Importantly, encoding models trained on English listeners with English-only embeddings generalized to brain activity in both Chinese and French listeners with remarkable predictive accuracy (Fig. 3C). We translated the story into 58 languages to map out the embedding space of a multilingual LM. We found clusters of languages that corresponded to their place in a comparative-linguistic family tree (Fig. 4B). Moreover, languages closer to a given subject's native language better predicted their brain activity (Fig. 4C). Overall, we demonstrated that both language models and brains encode conceptual meaning in a way that is partly agnostic to the specific forms of language used to convey such meaning. This raises an important theoretical question: why do different forms of language converge on similar conceptual representations?

First, all human languages must ultimately be couched in the "language" of neural computation. The diverse structures of language are likely encoded in a distributed fashion across populations of neurons (Saxena & Cunningham, 2019; Barack & Krakauer, 2021; Ebitz & Hayden, 2021), where individual neurons often exhibit "mixed selectivity" for a variety of different features (Fusi et al., 2016). The high-dimensional, vector-space representation of neural population codes provides a powerful format for capturing supramodal conceptual structure (Piantadosi et al., 2024) and generalizing to unique contexts (Hasson et al., 2020). LMs also rely on a continuous, high-dimensional embedding space to capture abstract, supramodal features across different languages and different modalities. There are many examples of this kind of supramodal representation, analogous to cross-language similarity (Honey et al., 2012), from different domains of neuroscience. Vision and language leverage shared, supramodal representations in the human brain (Quiroga et al., 2005; Mitchell et al., 2008; Popham et al., 2021; Tang et al., 2023; Wang et al., 2023) and in artificial neural networks (Goh et al., 2021; Radford et al., 2021; Huh et al., 2024). Reading and listening both rely on shared neural tuning (Deniz et al., 2019; Chen et al., 2024a). Speaking and listening, while two different processes, also recruit much the same neural systems (Stephens et al., 2010; Menenti et al., 2011; Silbert et al., 2014; Cai et al., 2023; Zada et al., 2024, 2025; Yamashita et al., 2025). At a global scale, the brain's connectivity is structured along a unimodal to transmodal gradient (Margulies et al., 2016). All of these examples suggest that the brain constructs abstract, supramodal representations, perhaps for the sake of efficiency in supporting flexible behavior (Dehaene & Cohen, 2007).

In LMs, it appears that language-specific features are rapidly projected into a more abstract feature space in relatively early layers, and these high-dimensional embeddings are only transformed back into more concrete word predictions in the final layers (Chang et al., 2022; Lindsey et al., 2025). Based on our current results, we propose that the human language system may follow a similar computational trend: particular linguistic inputs are rapidly transformed into an abstract, high-dimensional neural representational space capturing behaviorally-relevant meaning in the current context. We speculate that much of human thought may traverse this continuous, high-dimensional space of neural population activity, and only crystallize into the particular forms of language in moments of overt (or covert) speech.

Another reason why conceptual representations converge across languages hinges on how we learn and use language to communicate with one another. Humans are not born knowing the particular meaning of words in a particular language or culture, or how to use them effectively in particular contexts—we have to learn how to meaningfully *communicate* with others (Hasson et al., 2012; 2020; Fedorenko et al., 2024). For all of us, however, this learning process happens in a shared world. Different communities with different languages nonetheless interact with their environments in similar ways, with many similar social structures, as well as common behavioral needs and goals (Lupyan & Dale, 2016; Regier et al., 2016; Xu et al., 2016; San Roque et al., 2018). For example, many cultures include similar family structures, and may converge on some concepts of a "parent" across cultures. In the same way that different LMs trained on different languages nonetheless converge on shared structure, we suspect that human language learners also converge on conceptual representations keyed to the structures of our environment that are highly conserved across cultures.

Our findings do not provide support for the notion of universal properties of language. Linguistic phonology, morphology, syntax, and semantics are incredibly diverse, leaving little room for truly universal properties (Evans & Levinson, 2009). Language diversity stems from the community, culture, and context in which it is used (Majid et al., 2018; Thompson et al., 2020; Iordan et al., 2022). Although we found remarkable generalization in encoding across languages, the match is far from perfect, and we observed systematic gradations in cross-language generalization from more similar to less similar language families (Fig. 4). Here, we focused on commonalities rather than differences, partly due to the limitations of our data. First, we used only one book, which was written for children and young adults, thus limiting the complexity of the language used. Second, the book is professionally translated with the express goal of retaining the meaning of the original language (French). Both of these factors could reduce cross-language differences. Finally, our multilingual analyses were biased toward Indo-European languages, a common problem in the cognitive science and natural language processing literatures (Joshi et al., 2020; Blasi et al., 2022). In LMs, linguistic similarities between languages could also be confounded with geographic and cultural differences in the distribution of topics available in online text corpora. Future work could investigate more complex concepts and cross-linguistic differences, and include more language diversity.

Language is present in all human societies and is expressive enough to capture the incredible diversity of our experiences. This diversity, however, is undergirded by commonalities in the ways we interact with each other and with the world. By using language to communicate, human communities have encoded a great deal of the conceptual structure of our shared world into our different languages. Artificial language models, despite relying on very different neural machinery, have begun to key into the same shared structure. As these models become more embodied, more multimodal, and more interactive, we expect them to align even more closely with conceptual substructure of human experience (e.g., De Deyne, 2021; Xu et al., 2025). The success of these models reveals three sufficient ingredients for speakers to arrive at shared conceptual representations across languages: (1) a neural population code that is expressive enough to accommodate the context-rich structure of natural language; (2) a relatively simple statistical learning algorithm that drives an agent to encode useful structures of the environment; and (3) a learning environment of other speakers all interacting with a shared world. While the diverse *forms* of the world's languages are different on the surface level, the underlying *meaning* is shared across languages, reflecting our shared reality, and can guide us to better understand both our commonalities and our differences.

## Methods

*Participants*

We utilized the open dataset "Le Petit Prince fMRI Corpus," as released by Li and colleagues (2022) on OpenNeuro (Markiewicz et al., 2021) (https://openneuro.org/datasets/ds003643/versions/2.0.0). A total of 112 participants listened to the same audiobook in their native language: 49 English speakers (30 female, mean age 21.3), 35 Chinese speakers (15 female, mean age 19.9), and 28 French speakers (15 female, mean age 24.4). All participants were right-handed and self-identified as native speakers of their own language.

*Design and stimuli*

Three audiobook translations of the story Le Petit Prince (de Saint-Exupéry, 1943) were used for each language group. The audiobook lengths were 94 minutes, 99 minutes, and 97 minutes for the English, Chinese, and French translations, respectively. Li and colleagues (2022) provided word- and sentence-level transcripts for each audiobook. The English transcript consisted of 1,499 sentences (15,376 words); the Chinese transcript consisted of 1,577 sentences (16,009 words); and the French transcript consisted of 1,480 sentences (15,391 words). Each audiobook was split into nine ~10-minute segments such that each section starts and stops at the same meaningful boundary in all three languages.

Using the original transcripts as a starting point, we manually aligned sentences across all languages. This process involved either splitting long sentences or joining smaller sentences together so that each sentence has the same meaning across each language. Importantly, this process did not change the number of words in the story. It only changed which words correspond to which sentences. In total, we aligned 1,649 sentences across the three transcripts. Throughout the process, we leveraged translation services—e.g., Google Translate—to translate each French or Chinese sentence into English to help with the alignment. All transcripts are included in our GitHub repository.

*MRI acquisition*

Participants underwent neuroimaging while they listened to each ~10-minute section of the audiobook in their own native language across 9 separate runs. English participant imaging was acquired at Cornell University, United States; Chinese participant imaging was acquired at Jiangsu Normal University, China; and French participant imaging was acquired at NeuroSpin, France. English and Chinese participants were scanned using a 3T MRI GE Discovery MR750 scanner, while French participants were scanned with a 3T Siemens Magnetom Prisma Fit 230. All functional scans used a 2-second TR, a multi-echo pulse sequence (12.8 ms, 27.5 ms, 43 ms for English and Chinese; 10 ms, 24.46 ms, 38.92 ms for French), and a 3.75 mm$^3$ in-plane voxel resolution. For full details, we refer the reader to Li et al. (2022).

*Functional data preprocessing*

We used the preprocessed functional data that accompanied the "Le Petit Prince fMRI Corpus" dataset (Li et al., 2022). Li and colleagues preprocessed raw BOLD files using AFNI version 16. The pipeline was as follows. First, BOLD time series were corrected for slice-timing differences using AFNI *3dTshift* and despiked with *3dDespike*. Then, volumes were registered using *3dvolreg*. The functional data were nonlinearly aligned to the MNI template brain (*MNIColin27*). Finally, the data were preprocessed using multi-echo independent component analysis (ME-ICA). This procedure has been shown to remove noise associated with motion, physiology, and scanner artifacts (Kundu et al., 2012). Li et al. resampled the volumetric data to 2 mm$^3$ voxels using *3dresample*. However, we resampled the data back to its native 3.75 mm$^3$ resolution.

*Language model embedding extraction*

*BERT*
We used large language model activations to derive word and sentence embeddings from the transcripts. We used three unilingual BERT models (Devlin et al., 2018). These models are trained on text corpora of only one language. However, each model has the same underlying architecture (12 layers, 768 embedding dimensions) and objective function. We used the HuggingFace (Wolf et al., 2020) library for English BERT (*google-bert / bert-base-cased*), Chinese BERT (*google-bert / bert-base-chinese*), and French BERT (*dbmdz/bert-base-french-europeana-cased*). In addition, we used a multilingual BERT model that was trained on the 100 most used languages on Wikipedia (*google-bert / bert-base-multilingual-cased*). Importantly, training data for languages with larger corpora were subsampled, while training data for languages with smaller corpora were upsampled. We also created an untrained version of each BERT model where its weights were randomly initialized. This model's embeddings should not have any meaningful structure, but can still capture basic features and co-occurrence statistics in the stimulus itself (for example, word order or word frequency).

We extracted contextual word embeddings for each of the nine sections separately. Given a transcript, we first converted all text into tokens. Then, we defined inputs to the model in the form: <CLS> <context tokens> <SEP> <sentence tokens> <SEP>, where *sentence tokens* are the tokens of the current sentence from which we wish to extract embeddings and *context tokens* are the tokens before the sentence to fill the model context length up to 512 tokens. The special *CLS* and *SEP* tokens are included to match the training data used for BERT. Thus, for each token in <sentence tokens>, we extract the hidden states for each of the 12 layers, plus the input to the first layer, resulting in 13 total embeddings for each token. We repeat this process for each sentence within a section.

*Whisper*
We also extracted embeddings from the speech-to-text language model Whisper (Radford et al., 2022). We used *whisper-large-v3*, which is a multilingual model trained to transcribe speech to text in each of the 100 languages it supports. Whisper is an encoder–decoder transformer model that takes speech spectrograms as input to the encoder and text tokens as input to the decoder. The large model is composed of 32 encoder layers and 32 decoder layers. As we did with BERT, we extracted embeddings separately for each section, one sentence at a time. For each sentence, we extract a 30-second audio segment that ends with the current sentence's offset. At the same time, we extract text tokens that are associated with that segment, both in terms of the context tokens and current sentence tokens: <prefix tokens> <context tokens> <sentence tokens>. The *prefix tokens* are special tokens for whisper that include a task token (e.g., *transcribe*), a language token (e.g., French), and a start-of-sentence token. We input both the speech and its corresponding text to the encoder and decoder, respectively.

Whisper's encoder transforms the 30-second spectrogram input to 1,500 embeddings (a 50 Hz sampling rate) with 1,024 dimensions each. We map each word onset and duration to this sampling rate, and extract its corresponding embeddings, averaging into one embedding per word. We do this for each layer, retaining one embedding for each word per layer. Whisper's decoder outputs one embedding for each token, which we extract for the current sentence. For voxelwise encoding analyses, we used embeddings from the last layer of the encoder and embeddings from the middle layer of the decoder.

*Computing similarity between word embeddings*
We extracted contextual word embeddings for each word in the English, Chinese, and French transcripts from separate BERT models. To compute embedding similarity across languages, we averaged word embeddings within sentences to define a set of 1,649 sentence embeddings. So we ended up with 1,649 sentence embeddings for each unilingual BERT model per layer. Because each model's embeddings are structured in an arbitrary way (i.e., the embedding dimensions are not aligned), we opted to align them with a rigid rotation transformation before computing similarity. Specifically, for each layer, we used half of the embeddings to learn a Procrustes transformation from one language (e.g., Chinese) to another (e.g., English). Then, using the second half as a held-out test set, we rotated the held-out Chinese sentence embeddings and computed the correlation with the held-out English sentence embeddings. Finally, we averaged the correlations across the 768 embedding dimensions. We computed similarity for each pair of languages: English–Chinese, English–French, and Chinese–French.

For multilingual Whisper and BERT embeddings, we did not perform alignment via Procrustes. These models are forced to encode multiple languages into a unified embedding space, so we chose to compute similarity directly on the embeddings. Note that in the main figures, we report the similarity relative to a baseline. This baseline is derived from the same embedding extraction and analysis except that we use untrained, randomly initialized BERT models instead. This method serves as a strong baseline because everything remains constant except that the resulting embeddings would not be embedded in a structured geometry. We report the raw, un-subtracted, correlations in Fig S1.

*Whisper speech embedding phoneme classification*
We extracted phonetic pronunciations for each word in the transcripts from *ipa-dict* (https://github.com/open-dict-data/ipa-dict). All phonemes are based on the International Phonetic Alphabet (IPA). Across the transcripts of all three languages, we found a total of 63 unique phonemes (37 English, 40 Chinese, and 36 French). The intersection of all three languages was 19 phonemes. For each word in a transcript, we defined a 19-dimensional binary phonemic embedding vector indicating the presence of a phoneme with a 1.

We trained logistic regression classifiers to probe Whisper speech embeddings for phonemic categories in one language, then tested the classifier across languages. We fit separate binary classifiers for each of the 19 phonemes. We used five runs as our training set, and held out four runs to evaluate the models. We scored the models using adjusted balanced accuracy such that an accuracy of zero reflects chance performance. For each trained model, we tested its ability to decode phonemic categories across languages. For example, we fit a classifier to decode the presence of a shared phoneme from Whisper's English speech (encoder) embeddings. Then, we tested its ability to generalize and predict the presence of that same phoneme in the held-out runs using Whisper's Chinese speech embeddings and Whisper's French speech embeddings. We performed this analysis for Whisper's first and last encoder layers.

*Voxelwise encoding models*

To relate LLM word embeddings to the brain, we fit voxelwise encoding models to predict BOLD time series for each subject (Naselaris et al., 2011; Dupré La Tour et al., 2024). Encoding models take the form of linear regression, where the regressors are stimulus features (e.g., word embeddings in our case) and the dependent variables are the voxel time series. To account for the hemodynamic response, we copy and delay each regressor at four lags, from 1–4 TRs (2–8 seconds). This strategy effectively implements a finite impulse response. To account for the large number of regressors, due to the high dimensionality of LLM word embeddings, we use a ridge (L2) penalty to shrink the weights and reduce overfitting. We used the *RidgeCV* implementation from the *himalaya* library to search for the optimal penalty parameter in the training set (using nested cross-validation) across 20 numbers logarithmically spaced between $10^1$ and $10^{19}$ (Dupré La Tour et al., 2022). We used cross-validation when evaluating the models. Each model was fit on eight training runs, then tested on the held-out ninth run. Specifically, after training, we generate model-predicted time series based on the word embeddings in the held-out run, and then correlate the predicted time series with the subject's actual BOLD time series. We repeat this nine times, holding out one run for evaluation. Then, we average the correlation scores across all nine test sets to obtain an encoding performance correlation for each subject and voxel.

*Evaluating encoding models across languages*

In addition to evaluating each subject's encoding model on their own held-out BOLD responses, we evaluated the predictions against the average BOLD response of the other language groups. Averaging the BOLD time series among listeners of the same narrative isolates the shared, stimulus-driven neural responses (Hasson et al., 2004). This approach is typically used in intersubject correlation analyses without encoding models (Nastase et al., 2019). Building on this idea, several researchers have evaluated model-predicted time series *across* subjects to test for shared, *content*-driven features (Toneva et al., 2022; Zada et al., 2024). However, because our subjects listened to different forms of the same story, there is no precise temporal correspondence across languages. Thus, to compare neural time series across language groups, we downsampled the actual and predicted BOLD responses to the sentence level. We downsampled the time series by averaging TRs within a sentence's duration to 1,649 points for all 112 subjects.

We also defined the following confound time series variables at the sentence level: word rate, syllable/character rate, acoustic root mean squared (RMS) energy (via *librosa.feature.rms*), sudden acoustic changes (via *librosa.onset.onset_strength*), average framewise displacement, and the number of TRs averaged within each sentence. We regressed out these confounds from each voxel in the averaged, group BOLD response before correlating it with the model predictions.

Evaluating encoding models on the sentence level was a simplifying choice. For example, we could have learned an orthogonal transformation for stimulus features on the word-level (see Chen et al.,

2024b). Encoding weights learned from one language are abstracted away from the precise timing of that language and can be used to generate predictions for another language with different timing (de Varda et al., 2025). To further support our results, we evaluated encoding models for multilingual BERT on the TR level. For example, we trained encoding models to predict English subjects' BOLD from English mBERT embeddings. Then, using the same encoding model weights, we generate model predictions using Chinese (or French) mBERT embeddings. Although the embeddings are extracted using different transcripts, because the underlying model is the same, we expected the embeddings to be able to generalize. We found very similar encoding performance across languages to the sentence-level evaluation (Fig. S3).

*Translating the story*

We translated the English transcript to 55 languages that OpenAI officially supports—see Table S1 for a full list (we did not use Amharic or Somali because they were not in multilingual BERT's training data). We initiated GPT-4o with a sampling temperature of 0.5, and the developer-level prompt "Translate the English sentences into <language>.", where <language> was replaced with the target language name. Then, we sent a user message of the English sentence to translate, and received the translated sentence in the reply. We repeat this process for each sentence, keeping 3 English sentences and 3 translated sentences in history to provide context for the model. We manually inspected samples of the translated sentences to ensure there were no obvious errors, such as repeated words or clauses that can occur with greedy sampling.

We also collected language metadata from the Glottolog database 5.1 (Hammarström et al., 2024). We extracted each level of the language's position in the language family hierarchy.

*Software resources*

Our analyses relied on the following software: NiLearn (Abraham et al., 2014; RRID:SCR_001362), SciPy (Virtanen et al., 2020), scikit-learn (Pedregosa et al., 2011), Himalaya (Dupré La Tour et al., 2022), SurfPlot (Gale et al., 2021), and HuggingFace transformers (Wolf et al., 2020).

## Acknowledgments

This work was supported by the National Institutes of Health grants R01DC022534 (U.H.). We thank everyone involved with collecting and sharing the "Le Petit Prince" multilingual naturalistic fMRI corpus. We also thank members of the Hasson lab and the NeuroML lab, led by Meenakshi Khosla, for their feedback and discussions. Finally, thanks to Haocheng Wang for help with Chinese translation.

*Author Contributions*

Z.Z., S.A.N., and U.H. conceptualized the project. Z.Z. performed the analyses, wrote the software, visualized the results, and wrote the initial draft. S.A.N. and U.H. edited and revised the manuscript.

*Competing interests*

Authors declare that they have no competing interests.

*Data and materials availability*

Source code and materials are available at https://github.com/zaidzada/crosslingual-convergence. Data is available at https://openneuro.org/datasets/ds003643/versions/2.0.0.

# Supplementary Materials

**Brains and language models converge on a shared conceptual space across different languages**

Zaid Zada[1*], Samuel A. Nastase[1], Jixing Li[2], Uri Hasson[1]

[1] Department of Psychology and Neuroscience Institute, Princeton University, New Jersey, 08544, US.
[2] Department of Linguistics and Translation, City University of Hong Kong
* Corresponding author. Email: zzada@princeton.edu

Includes figures S1–S4, and table S1.

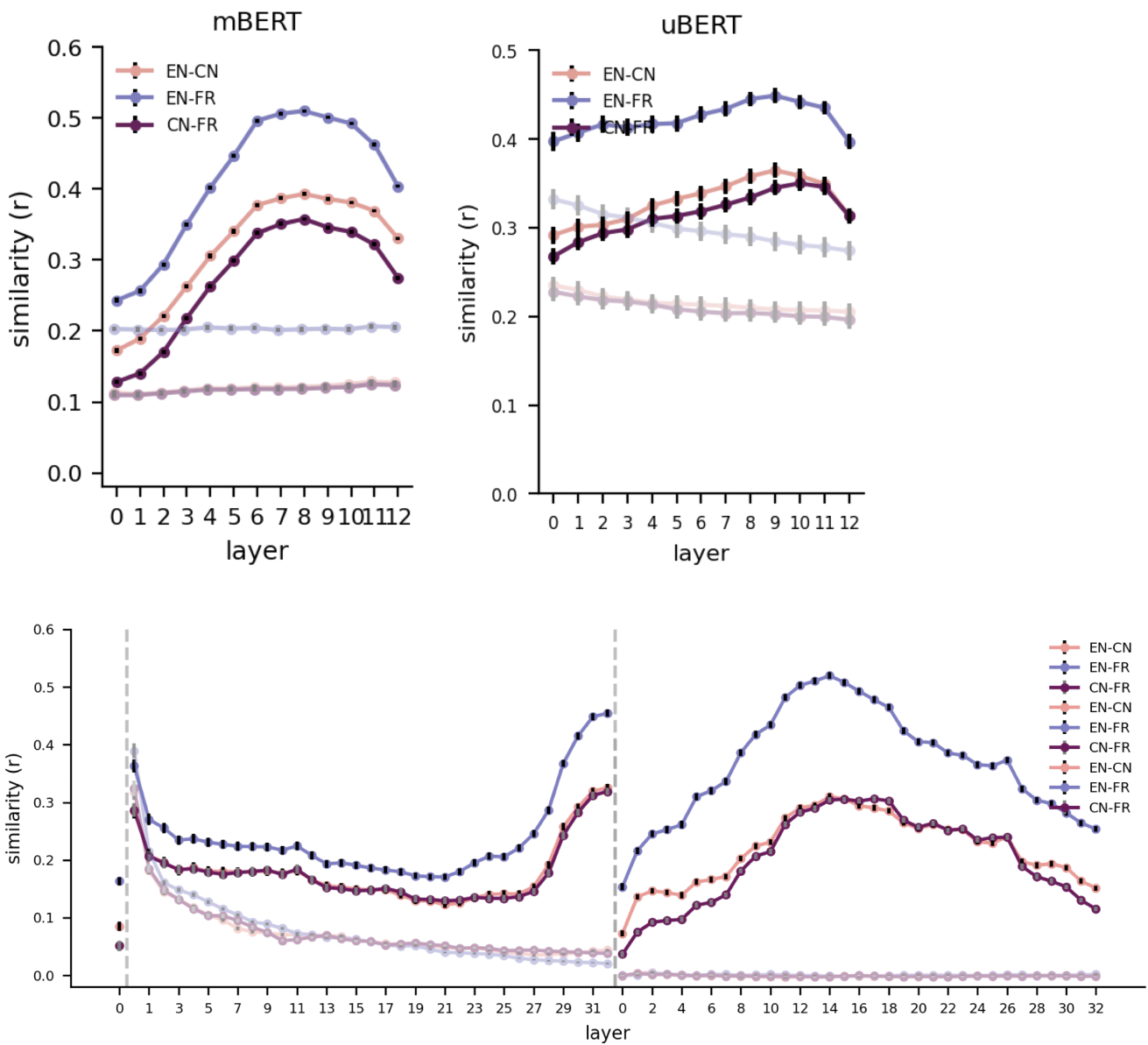

**Fig. S1. Trained and untrained sentence-embedding similarities without normalizing to a baseline.** The correlations between sentence embeddings across different language pairs is presented relative to a baseline in the main figures. Here, we show the raw correlations as well as the baseline which we subtracted—which is derived using the exact same procedure but with untrained models—randomly initialized parameters—for unilingual BERT, multilingual BERT, and Whisper.

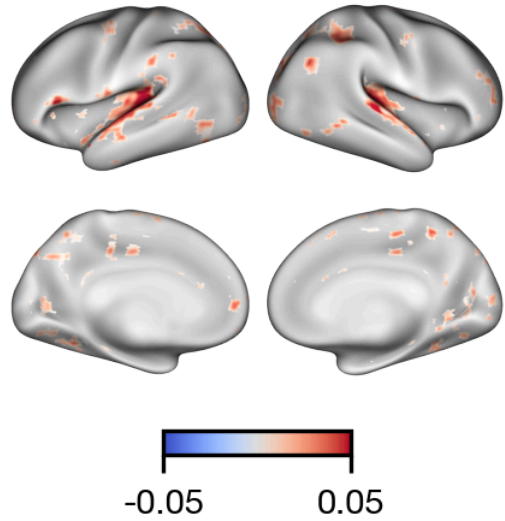

**Fig. S2. Difference map between within- and across-language encoding performance.** We statistically tested the difference between within- and across-language encoding performance (as seen in Fig. 3B, 3C). P-values were corrected for multiple comparisons using FDR ($p_{FDR} < .05$).

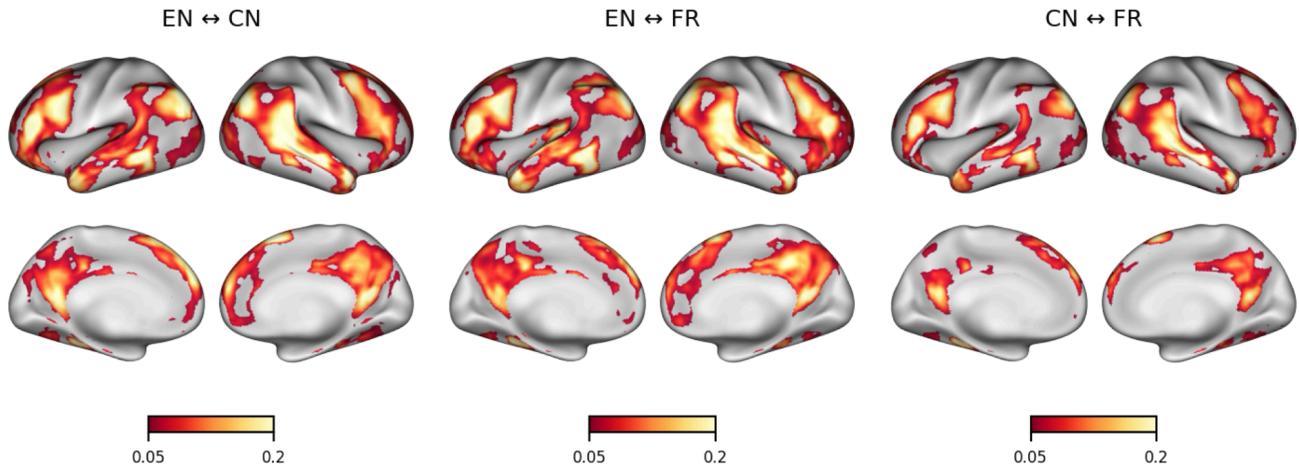

**Fig. S3. Alternative strategy for calculating cross-language encoding generalization using multilingual BERT.** In the main analyses, we chose to downsample all time series to the sentence level to compare actual and model-predicted brain activity across languages. However, there is an alternative method that can be used on the original, TR-level, time series. Here, we use a trained encoding model on one language (e.g., English mBERT embeddings → English listener), and use the same weights to produce a model predicted brain response using mBERT embeddings from another language (e.g., Chinese mBERT embeddings) and test it on that language's group average brain activity. The result closely replicates our sentence-level analysis findings.

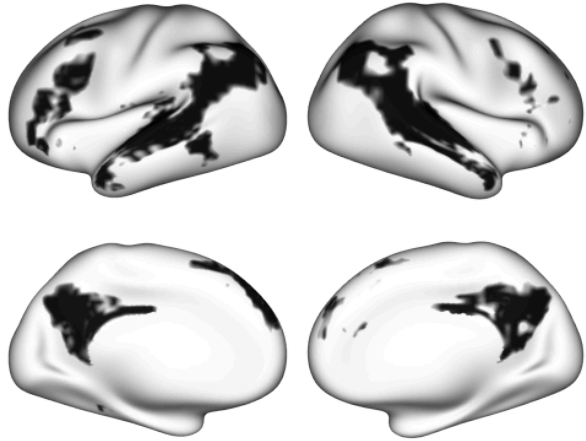

**Fig. S4. Brain mask used to select voxels for summarizing whole-brain encoding performance.**
We defined a brain mask based on intersubject correlation (ISC) to select voxels that exhibited stimulus-dependent brain responses shared across subjects. We used this mask to summarize encoding performance across the entire brain, for example in Fig. 4.

| code | name | family | sub-family | code | name | family | sub-family |
|---|---|---|---|---|---|---|---|
| ARA | Arabic | Afro-Asiatic | Semitic | DUT | Dutch | Indo-European | Germanic |
| SWA | Swahili | Atlantic-Congo | Volta-Congo | EN | English | Indo-European | Germanic |
| VIE | Vietnamese | Austroasiatic | Vietic | GRE | Greek | Indo-European | Graeco-Phrygian |
| TGL | Tagalog | Austronesian | Greater Central Philippine | BEN | Bengali | Indo-European | Indo-Iranian |
| MAY | Malay | Austronesian | Malayo-Chamic | PAN | Punjabi | Indo-European | Indo-Iranian |
| IND | Indonesian | Austronesian | Malayo-Chamic | MAR | Marathi | Indo-European | Indo-Iranian |
| KAN | Kannada | Dravidian | South Dravidian I | GUJ | Gujarati | Indo-European | Indo-Iranian |
| MAL | Malayalam | Dravidian | South Dravidian I | HIN | Hindi | Indo-European | Indo-Iranian |
| TAM | Tamil | Dravidian | South Dravidian I | URD | Urdu | Indo-European | Indo-Iranian |
| TEL | Telugu | Dravidian | South Dravidian II | PER | Persian | Indo-European | Indo-Iranian |
| ARM | Armenian | Indo-European | Armenic | RUM | Romanian | Indo-European | Italic |
| LAV | Latvian | Indo-European | Balto-Slavic | ITA | Italian | Indo-European | Italic |
| LIT | Lithuanian | Indo-European | Balto-Slavic | FR | French | Indo-European | Italic |
| UKR | Ukrainian | Indo-European | Balto-Slavic | SPA | Spanish | Indo-European | Italic |
| RUS | Russian | Indo-European | Balto-Slavic | POR | Portuguese | Indo-European | Italic |
| BUL | Bulgarian | Indo-European | Balto-Slavic | CAT | Catalan | Indo-European | Italic |
| MAC | Macedonian | Indo-European | Balto-Slavic | ALB | Albanian | Indo-European | Classical Indo-European |
| BOS | Bosnian | Indo-European | Balto-Slavic | JPN | Japanese | Japonic | Japanesic |
| HRV | Croatian | Indo-European | Balto-Slavic | GEO | Georgian | Kartvelian | Georgian-Zan |
| SRP | Serbian | Indo-European | Balto-Slavic | KOR | Korean | Koreanic | Koreanic |
| SLV | Slovenian | Indo-European | Balto-Slavic | MON | Mongolian | Mongolic-Khitan | Mongolic |
| CZE | Czech | Indo-European | Balto-Slavic | BUR | Burmese | Sino-Tibetan | Burmo-Qiangic |
| SLO | Slovak | Indo-European | Balto-Slavic | CN | Chinese | Sino-Tibetan | Sinitic |
| POL | Polish | Indo-European | Balto-Slavic | THA | Thai | Tai-Kadai | Kam-Tai |
| SWE | Swedish | Indo-European | Germanic | KAZ | Kazakh | Turkic | Kipchak-Turkestan |
| DAN | Danish | Indo-European | Germanic | TUR | Turkish | Turkic | Oghuz |
| ISL | Icelandic | Indo-European | Germanic | EST | Estonian | Uralic | Finnic |
| NOR | Norwegian | Indo-European | Germanic | FIN | Finnish | Uralic | Finnic |
| GER | German | Indo-European | Germanic | HUN | Hungarian | Uralic | Hungaric |

**Table S1. List of all languages used and their location in a language family hierarchy.** We translated the English story transcript into languages that are supported by the GPT-4o model which we used for translation (https://help.openai.com/en/articles/8357869-how-to-change-your-language-setting-in-chatgpt), and languages that were used for training multilingual BERT (https://github.com/google-research/bert/blob/master/multilingual.md). The language hierarchy is based on the Glottolog database (Hammarström et al., 2024). The family column indicates the top-level (root node) family for that language, whereas the sub-family column indicates the next top-level parent node that divides the languages used.